\documentclass[12pt]{article}
\usepackage{graphicx}
\usepackage{natbib}
\bibpunct{(}{)}{;}{a}{}{,}
\usepackage{txfonts}
\def\p0943{PSR B0943$+$10}
\begin{document}

\title{The issue of aliasing in \p0943 \\ I Limitations of the physical model}
\author{M. Vivekanand\footnote{vivek@ncra.tifr.res.in} \\
	National Center for Radio Astrophysics, TIFR, \\
	Pune University Campus, P. O. Box 3, \\
	Ganeshkhind, Pune 411007, India.}
\maketitle

\noindent {\bf Abstract}:
\cite{DR1999, DR2001} claim that the frequency of the very narrow feature, in the
spectrum of radio flux variations of \p0943, is an alias of its actual value. For 
this, they need to interpret the narrow feature as a very stable pattern of 
drifting sub pulses on the polar cap of this pulsar, within the framework of the 
\citet{RS1975} model. This paper discusses several factors within the same 
framework, that may perturb a very steady pattern of drifting sub pulses, leading 
to a broadening of the spectral feature. One should be cautious in (1) identifying 
such very narrow spectral features with a very stable pattern of drifting sub pulses,
and (2) use this picture to resolve the issue of aliasing in \p0943.

\section{Introduction}

The rotation powered radio pulsar \p0943 has a very narrow feature in the longitude resolved 
spectrum of intensity fluctuations; its $Q$, defined as its central frequency divided by its 
width, is relatively high (\citealt{TH1971}; \citealt{BRC1975}; \citealt{SO1975}). Recently 
\citeauthor{DR1999} (\citeyear{DR1999}, \citeyear{DR2001}; henceforth DR1999 and DR2001) put 
its $Q$ at $ >> 500$, and also claim that this very narrow spectral feature, occurring at 
0.465 cycles per pulsar period [cpp], is an alias of the actual value 0.535 [cpp]. \cite{DB1973} 
noted such a general possibility in radio pulsars and claimed, based on the phase information 
in the fluctuation spectrum, that PSR B2303+30 has an aliased spectral feature. This was 
contested by \cite{SO1975}, who state on their page 326 that ``even a phase analysis, contrary 
to Backer's statements (\citealt{DB1973}), is unable to decide between the two possibilities'', 
viz., whether the spectral feature is aliased or not. Indeed, in a later paper Backer did not 
repeat such a claim for \p0943, whose fluctuation spectrum is similar to that of PSR B2303+30 
(\citealt{BRC1975}); he quoted the the true value as 0.465 [cpp].  However, In the view of 
DR2001, \cite{SO1975} ``came to the wrong conclusion''.

Like earlier workers, DR2001 associate this high $Q$ spectral feature with the well known 
phenomenon of systematically drifting sub pulses in radio pulsars, and offer three arguments, 
involving signal processing of position modulated pulses, which are by themselves not 
sufficient to resolve the aliasing issue. They need to make the crucial assumption that the 
\cite{RS1975} model (henceforth RS1975) applies to the data of \p0943. Only then are they 
able to interpret the high $Q$ feature as arising on account of 20 uniformly spaced sparks 
on the polar cap, rotating very stably round the magnetic axis, once every $\approx$ 37.35 
pulsar periods ($\approx 41$ s). Only after this are they are able to study the issue of 
aliasing and claim to have resolved it. For a brief and clear summary of the crucial argument 
of DR2001 regarding aliasing, see the second last para on page 739 of \citet{ES2002}, and the 
corresponding footnote.

This is the first paper in the series that critically analyzes the DR1999 and DR2001 papers 
regarding the issue of aliasing. It argues that they are not justified in using this model 
as an essential input to resolve the issue of aliasing in \p0943; there are several factors 
within the framework of that model, that are capable of perturbing very steadily drifting 
sub pulses, thereby reducing the $Q$ of the corresponding spectral feature to values much 
smaller than 500. The above factors are not important for the canonical drifting pulsars, 
in which the $Q$ is not higher than $\approx$ 10. In this paper the RS1975 model also 
implies the subsequent modifications made to it by Ruderman and colleagues, while retaining 
its essential features (\citeauthor{CR1977} \citeyear{CR1977}, \citeyear{CR1980}; henceforth 
CR1977 and CR1980). 

\begin{figure}
\includegraphics[width=14.0cm]{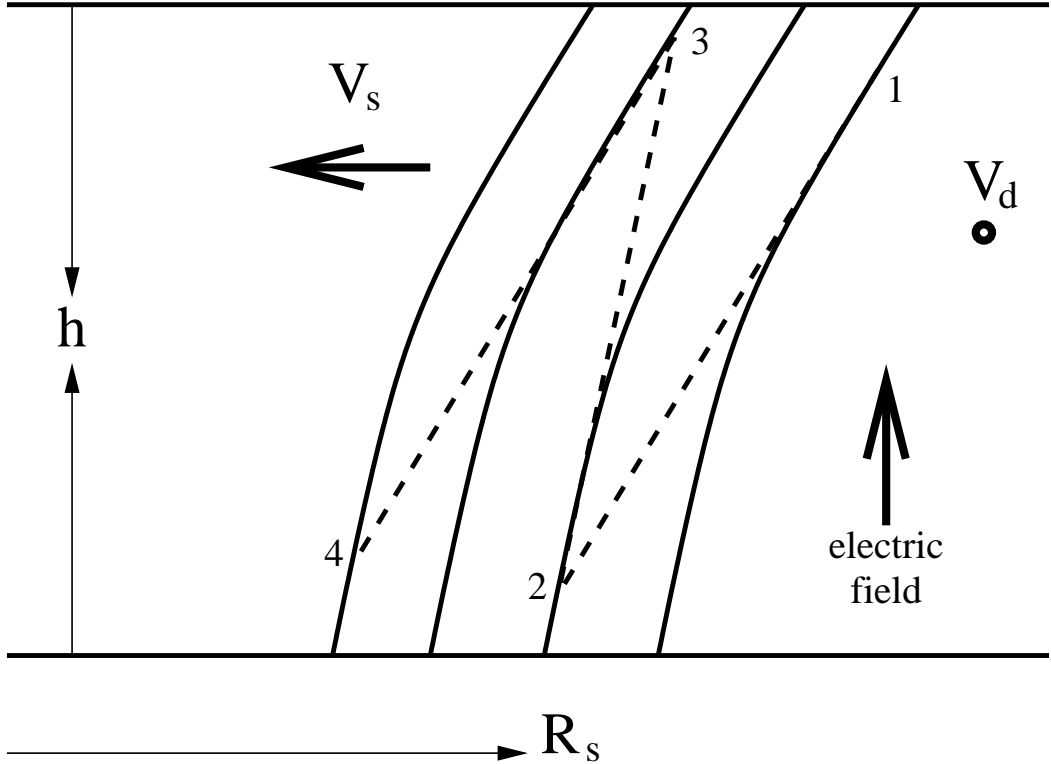}
\caption{
	Similar to fig.~1 of CR1977. A high energy photon initiates a spark at position 
	1, by forming an $e^+-e^-$ pair due to the Erber process (see RS1975), in a 
	vacuum polar gap of height $h$. The $e^-$ moves along the curved magnetic 
	$\mathbf{B}$ field line, accelerated down wards by the electric field 
	$\mathbf{E}$, and emits a high energy photon due to curvature radiation. This 
	travels tangentially to the original $\mathbf{B}$ field line and creates 
	another $e^+-e^-$ pair at position 2. The $e^+$ of this pair creates another 
	pair at position 3, and so on. Thus the spark shifts radially to wards the 
	magnetic axis with speed $V_s$, while simultaneously drifting tangentially with 
	speed $V_d$ in the crossed $\mathbf{E}$ and $\mathbf{B}$ fields. $R_s$ is the 
	radial distance of the spark from the magnetic axis (which is not shown here).
	}
\label{fig1}
\end{figure}

RS1975 derive the characteristic period $P_3$ for a sub pulse position to recur as
\begin{equation}
P_3 = \frac{1}{N} \times \frac{2 \pi R_s}{V_d} \times \frac{1}{P} \ \ \mathrm{periods,}
\ \ \ V_d = c \left \vert \frac{\mathbf{E} \times \mathbf{B}}{\vert \mathbf{B} \vert^2}
\right \vert ,
\end{equation}
\noindent
where $R_s$ is the radial distance of the spark from the magnetic axis, $N$ is the number 
of sparks laid out uniformly on the circumference $2 \pi R_s$, $P$ is the pulsar period, 
and $V_d$ is the tangential speed of the spark, due to drift in crossed electric $\mathbf{E}$ 
and magnetic $\mathbf{B}$ fields in the vacuum polar gap (fig.~\ref{fig1}); $c$ is the speed 
of light. The frequency of the feature in the longitude resolved fluctuation spectrum is 
then $1/P_3$ [cpp]; this paper discusses the stability of this quantity.

Let the parameters $R_s$, $V_d$ and $N$ of eq. (1) vary independently with 
time, due to reasons described below, with standard deviations $\sigma_{R_s}$, 
$\sigma_{V_d}$, and $\sigma_{N}$, respectively. Then the expected $Q$ of the 
spectral feature is given by 
\begin{equation}
Q \approx \left [ \left ( \frac{\sigma_{R_s}}{R_s} \right )^2 + \left ( 
	\frac{\sigma_{V_d}}{V_d} \right )^2 + \left ( \frac{\sigma_{N}}{N} 
	\right )^2 \right ]^{-1/2}
\end{equation}
\section{Radial drift of spark in RS1975}

First, strict RS1975 scenario is assumed; i.e., the vacuum gap breaks down due pair 
production in the gap $\mathbf{B}$ field, and not due to pair production in the 
Coulomb $\mathbf{E}$ field of Fe ions (CR1977). Then the velocity of radial shift 
(or spread) of the spark is given by (CR1980)
\begin{equation}
V_s \approx \frac{1}{2} \frac{c h}{S}
\end{equation}
\noindent
where $h$ is the height of the vacuum gap and $S$ is the radius of curvature of the 
$\mathbf{B}$ field lines. 

For the required $\mathbf{B}$ field line curvature of $S \approx 10^6$ cm (RS1975, 
CR1977, CR1980), the time to cross the polar cap ($\approx 10^4$ cm) is $\approx 
100$ micro seconds ($\mu$s); see CR1980 for details. Therefore 
in the duration of a spark, which RS1975 estimate to be 1 $\mu$s to 10 $\mu$s, 
which also tallies with the micro structure time scale expected in a pulsar 
with period 1.0977 s (\p0943), one can expect the spark to have traveled $\approx$ 
1 meter (m) to 10 m. Thus one can expect $\sigma_{R_s}$ to be $>> 1$ m due
to this process.

Now, DR1999 invoke 20 sparks, each of size $\approx$ 20 m, uniformly spaced apart 
from each other by $\approx$ 45 m, on a circumference of radius $\approx$ 145 m on
the polar cap. Thus the spark might have moved radially in wards by a fraction 
$\ge 1/20$ of its size during its own development, leading to  $R_s / 
\sigma_{R_s} \le 145$, implying that the $Q$ of eq. (2) might be limited to the 
value $\le 145$.

CR1977 discuss this issue and propose two solutions for dealing with it, 
which is a departure from the strict RS1975 model. Their proposals probably do
not work for high $Q$ drifting as assumed in \p0943.

First they propose that the $\mathbf{B}$ field line geometry could be sufficiently 
complicated to stop such a radial shift; but this might also make the canonical
tangential drift non-uniform, and once again decrease the $Q$ of the spectral 
feature due to drifting. CR1977 also discuss another fundamental problem 
with this scenario.

Second, they propose that the relativistic electrons that penetrate the surface 
of the neutron star might initiate a particle shower which might eventually lead 
to a high energy photon reentering the vacuum gap, to produce $e^+ - e^-$ pairs. 
This allows the spark to be initiated on the original $\mathbf{B}$ field line in 
the vacuum gap, because one has now diluted the role of curvature radiation in
the gap break down process. This might cause two problems for very stable (high 
$Q$) drifting: (a) unless this process is highly probable and occurs uniformly
in space and time, it might cause erratic sparking in the vacuum gap, which might 
lead to a reduced $Q$ of the drifting spectral feature, due to, say, random 
variations in shape, size and position of sub pulses; for example, small 
variations in the angle of entry into the gap of these high energy photons may 
change the position of the centroid of the spark; (b) unless the curvature 
related pair production ceases completely, one can not avoid altogether the radial 
shifting of the spark. This may lead to an effective widening of the spark area 
along the radial direction, if not to an outright shifting of the spark, which 
may also reduce the $Q$.

CR1977 also propose an even more radical departure from the strict RS1975 model. 
They propose that alternatively (to the Erber process) the gap can break down 
by $e^+ - e^-$ pair production in the Coulomb field of Fe ions, which are released 
into the gap by the hot polar cap surface, by thermal x-ray photons from the same 
hot polar cap. They call this the Lorentz boosted pair production. The relativistic 
electrons returning to the polar cap surface maintain it hot. In this scenario also 
there are two problems. First, it is not clear what will be the final space 
distribution of the charges making up the spark -- it may be much wider than 
$\approx 20$ m, and its centroid may differ from spark to spark by more than 1 m 
in position. Second, even if a narrow spark was drifting, and it carried underneath 
it the local hot spot, the cooling time for the polar cap surface is fairly long, 
of the order of $\approx 0.2$ s to even 10 s (CR1980). Therefore the spark may not 
drift as one observes, but may merely widen in the tangential direction. The 
non-uniformity of this widening may once again causing problems for high $Q$ 
drifting spectral features.

\section{Change in drift speed of spark}

\vfill \eject

So far the drift speed $V_d$ was assumed to be unvarying. However in the strict RS1975 
model, it may also depend upon the radial distance $R_s$ (see eq. (30) of RS1975 and 
the subsequent discussion). Thus a perturbation in $R_s$ could also lead to a 
perturbation in $V_d$, further reducing the observable $Q$ in eq. (2).

Apart from the above, there appear to be very few perturbations that directly affect 
the drift speed $V_d$, at least to the levels discussed here. However, the curvature 
drift plasma process might set an upper limit to the observable $Q$. This arises due 
to the motion of charges along a curved path. The corresponding drift speed (in the 
tangential direction) is then given by (\citealt{LL1975}; \citealt{FFC1984})
\begin{equation}
V_c = \frac{\gamma}{S \omega_B} \left ( v^2_\parallel + \frac{1}{2} v^2_\perp \right ) 
\left \vert \hat{\mathbf{S}} \times \hat{\mathbf{B}} \right \vert,
\end{equation}
\noindent
where $\gamma = 1 / \sqrt{1 - v^2/c^2}$, $v_\parallel$ and $v_\perp$ are the speeds 
of the charges parallel and perpendicular to the $\mathbf{B}$ field, $\omega_B = e 
\vert \mathbf{B} \vert / m_e c$ is the cyclotron frequency, and $\hat{\mathbf{S}}$ 
and $\hat{\mathbf{B}}$ are unit vectors normal and parallel to the $\mathbf{B}$ 
field line, respectively. Setting $v_\parallel \approx c$, $v_\perp \approx 0$, 
$\omega_B \approx 10^{19}$ Hz, $S \approx 10^6$ cm, and $\gamma \approx 10^2$ to 
$10^3$ (RS1975, CR1977, CR1980), one obtains $V_c \approx 10^{-4}$ m/s to $10^{-3}$ m/s. 
Comparing this with the $V_d = 2 \pi R_s / 41 \approx 22$ m/s speed of sparks required 
by DR1999, 
the perturbation to $V_d$ can be estimated to be $\sigma_{V_d} / V_d$ of $>> 0.5 
\times 10^{-5}$ to $0.5 \times 10^{-4}$. Currently the latter number would imply a 
$Q$ that is a comfortable factor of $\approx 40$ higher than the $Q >> 500$ claimed 
by DR2001. So in principle it might appear that one can ignore this perturbation to 
$Q$ in eq. (2).

However, DR2001 claim that ``longer fast Fourier transforms yield a progressively 
narrow feature'' in their section 3; and they have used a data of length 816 periods
only. Even then they claim that the spectral feature ``is only partly resolved''.
One should worry about the possibility that future (and longer) 
observations might bring the observed $Q$ of \p0943 uncomfortably close to the 
above upper limit.  Of course, one should eventually do a full relativistic quantum 
calculation of eq. (4), and worry about the possibility that future theoretical 
inputs might also change the above upper limits.

Furthermore, the above drift is charge dependent, i.e., the spark might get 
charge separated. This may not only widen the spark as it develops, it may also 
have very important consequences for the simultaneous existence of several 
isolated and independent sparks on the polar cap, as required by DR1999.

\section{Change in number of sparks}

In eq. (2) a random change of number of sparks $N$ by even $\pm 1$ causes the $Q$ 
to be $\approx 20$, much less than the $>> 500$ claimed by DR2001. Therefore any 
change in $N$ is not allowed in the scenario of DR1999; which is why they claim 
that in \p0943 ``20 sub beams rotate almost rigidly, maintaining their number and 
spacing despite perturbations ...'', and that ``the 20-fold pattern is always 
maintained strongly as a stable configuration'' (on the right side of page 1010 
in their paper). Now, there are two important elements of the physics of pulsars 
that, together, might not allow the above requirement of DR1999.

First, the polar cap surface is supposed to be a good electrical conductor; this
is a necessary requirement for the basic electrodynamics of the pulsar 
\citep{GJ1969}. Even if the strong $\mathbf{B}$ field makes the transverse 
conductivity at the surface small, there must be a capacitive conductivity (i.e.,
conductivity with a time delay caused by, say, returning currents) near the polar 
cap surface, for the basic electrodynamics to hold true.

Second, the gap break down process is supposed to be an extremely sensitive one.
RS1975 state in their section 5 that ``the extreme sensitivity of the discharge 
growth to small variations in the gap thickness may be reflected in the erratic 
nature of pulsar sub pulses''. Later modifications to the RS1975 model dilute 
this sensitivity, since RS1975 assumed a gap potential difference that varies 
as $h^2$, while space charge limited flow in the gap makes it vary only as $h$ 
(eqs. (5) and (6) of CR1980). However, these later modifications also made the gap 
potential difference very sensitive to the thermionic emission from the polar cap 
surface, which depends exponentially upon the surface temperature, which in turn 
depends upon the return current heating the surface (CR1980). CR1980 state 
at the end of their section 3 that ``This extreme sensitivity of the potential 
drop to the surface temperature makes the gap thermostatically regulating ...''.

Given these two aspects of pulsar polar gap physics, the simultaneous existence of
$\approx 20$ stable sparks on the polar cap surface may not be assured. The spark
growing the fastest on the polar cap might reduce the overall gap potential 
difference, thereby inhibiting the growth of other sparks; at least one 
observational paper has claimed to have seen competition in intensity between 
drifting sub pulses (of PSR B0031$-$07; \citealt{VJ1999}). Observations certainly 
allow for 2 or 3 simultaneous sparks on the polar cap (see the earlier references; 
also see \cite{VJ1999}). However, to the best of the knowledge of this author, no 
one has seen as many as $>> 5$ simultaneous sub pulses within a period.

In view of the above, 
the existence of $\approx$ 20 simultaneous sparks on the polar cap is highly 
suspect, at least within the current RS1975 framework.

\section{Thermostatic regulation of gaps}

CR1977 state in their abstract that ``A potential drop above a polar cap may
arise from (A) the inertia of freely released ions in a space-charge limited flow, 
or from (B) surface binding of ions leading to gap formation''. CR1980
develop this idea, and claim that the gap potential will be, in general, intermediate
between the value for a true vacuum gap and the value for a space charge limited flow.
Such a situation can easily break into oscillations -- a perturbation leading to an
increased surface temperature will cause higher ionic current from the polar cap, 
resulting in a higher space charge and therefore a lower gap potential. This would 
reduce the (return) heating current onto the polar cap, thus cooling it, and reducing
the ionic current. And vice versa when the perturbation decreases the surface 
temperature. From eq. (14) of CR1980, one can estimate the (cooling) time for the 
polar cap temperature to decrease by a factor of 10 to be $\approx 0.2$ s; from their 
eq. (16) the (heating) time for the polar cap temperature to increase by a factor of 
10 is $\approx 0.008$ s. However CR1980 state that the actual times may be larger due 
to the approximations made.

Therefore any spot on the polar cap, across which a spark has drifted, might have 
in general a quasi-periodic temperature variation, with a rise time of a few times 
$\approx 0.008$ s and a fall time of a few times $\approx 0.2$ s; the period of this
variation will be the time for a spark to recur at the same spot, which is $P_3$ 
pulsar periods. How periodic or quasi-periodic will depend upon the actual period and 
the rise and fall time profiles, and their variation. The above time scales will change 
if the threshold is set not to increase or decrease of temperature by factor 10, but 
by some other factor.

Now the gap potential is exponentially sensitive to this polar cap surface temperature 
variation (CR1980). So one may expect to see a corresponding quasi-periodic intensity 
modulation of the drifting sub pulse. In the spectrum of intensity fluctuations, this 
will be evident as side bands beating with the drifting spectral feature; their width 
and spacing will depend upon $P_3$ and $Q$ of the modulating signal. Indeed, DR1999 
claim that in \p0943 the drifting feature at period $P_3 \approx 1.87$ periods is 
amplitude modulated by a secondary periodicity of period $\approx 37.35$ pulsar periods 
($\approx$ 41.0 s). However, this amplitude modulating signal also appears to have a 
very high $Q$, of the order of 500 to 1000. Whatever be the physical reason for such 
amplitude modulation, one should ensure theoretically that such an observation is 
possible; i.e., the quasi-periodic temperature variation discussed above, which must 
be surely operative in the scenario of CR1980, must have a very high $Q$, at least 
higher than 500 to 1000. If not, then a low $Q$ quasi-periodic temperature variation 
will broaden any narrow drifting feature in the fluctuation spectrum.

From the theoretical viewpoint, a problem that the scenario of DR1999 might pose, 
is that any spot on the circumference $2 \pi R_s$ is heated 20 times faster than 
anticipated earlier, i.e., once every 1.87 pulsar periods. This might keep the 
average polar cap temperature hotter than earlier anticipated; please note that
the heating plus cooling timescale may be comparable to this periodicity. Then one 
might expect to see a significant amount of pulse nulling in the 816 periods of 
\p0943, whereas not even one pulse is observed to be nulled (DR2001).

\section {Conclusion}

This paper examined the issue within the frame work of the RS1975 model, along
with modifications made by CR1977 and CR1980. However, It is well known that 
the circular symmetry for the polar gap assumed in RS1975 is an idealization.
A realistic pulsar, having mis aligned rotation and the magnetic axis, has 
a north-south electric field on the polar cap \citep{FCM1991}. This causes the 
sub pulse drift to be mainly in the east-west direction on the polar cap, 
instead of being circular (fig.~2b of \citealt{MR1976}). This may modify 
significantly the conclusions of DR1999, where some sort of circular symmetry 
is essential.

This paper is not a criticism of the RS1975 model. It merely disagrees with its 
use, given its limitations, as an essential input to resolving the issue of 
aliasing of the high $Q$ spectral feature in \p0943, even if that can actually 
be associated with the drifting sub pulse phenomenon. If such an association is
actually invalid, then as to what actually is this spectral feature is a 
speculation beyond the scope of this work.

\noindent {\bf Acknowledgments}:
This research has made use of NASA's Astrophysics Data System (ADS) Bibliographic 
Services.

\vfill
\eject
\end{document}